\documentclass{parcfd}
\pdfoutput=1

\usepackage{cite}
\usepackage[pdfpagelabels]{hyperref}
\usepackage{graphicx}
\usepackage{amsmath}

\usepackage{color}
\definecolor{grey}{RGB}{170,170,170}
\usepackage{listings}

\usepackage{pgfplots}
\usepackage{pgfplotstable}

\pgfplotsset{compat=1.8}
\usepgfplotslibrary{statistics}
\usepackage{xurl}

\usepackage{float}
\restylefloat{table}

\usepackage{wrapfig}

\title{Verificarlo CI: continuous integration for numerical optimization and debugging}

\author{Aurélien Delval$^{1,2}$,
François Coppens$^{1}$,
Eric Petit$^{3}$, 
Roman Iakymchuk$^{4}$,\\
Pablo de Oliveira Castro$^{1}$}

\heading{A. Delval, F. Coppens, E. Petit, R. Iakymchuk, P. de Oliveira Castro}

\address{$^{1}$Université Paris-Saclay, UVSQ, LI-PaRAD, France.\\
Email: \{aurelien.delval, francois.coppens, pablo.oliveira\}@uvsq.fr\\ 
$^{2}$SiPearl, France. Email: aurelien.delval@sipearl.com\\
$^{3}$Intel Corporation, USA. Email: eric.petit@intel.com\\
$^{4}$Ume\aa{} University and Uppsala University, Sweden. Email: riakymch@cs.umu.se}

\abstract{Floating-point accuracy is an important concern when developing numerical simulations or other compute-intensive codes. Tracking the introduction of numerical regression is often delayed until it provokes unexpected bug for the end-user.  In this paper, we introduce Verificarlo CI, a continuous integration workflow for the numerical optimization and debugging of a code over the course of its development. 
We demonstrate applicability of Verificarlo CI on two test-case applications.}

\begin{document}
\section{Introduction}

Despite Floating-point (FP) accuracy being a known issue~\cite{goldberg}, modern tools for software development do not provide automated numerical accuracy regression tests.  To fill this need, we propose Verificarlo CI (Continuous Integration).
GitHub and GitLab are popular platforms for developing software, and both have features for CI. CI services are triggered on specific events, such as merging a pull request. Verificarlo CI is designed to be integrated with them, but it can also be used with custom workflows.

To facilitate its adoption, Verificarlo CI has been designed to be easy and fast to deploy, while still being flexible enough to be relevant for most applications. We provide the user with a simple API to insert FP probes in their tests, execute them with Verificarlo, setup CI Actions, and finally access and interpret the results.

Finally, we demonstrate Verificarlo CI on two use-cases: exploring reduced mixed-precision in the Nekbone CFD proxy application; tracking numerical bugs during the development of a the Quantum Monte Carlo Chemistry Kernel library (QMCkl).

\section{Verificarlo CI for numerical correctness}
\label{vfc_ci_architecture}

Verificarlo~\cite{verificarlo} is a tool based on the LLVM compiler framework modifying at compilation each floating point operation with custom operators. After compilation, the program can be linked against various backends to explore FP related issues and optimizations. The latest version of Verificarlo fully supports OpenMP and MPI parallel programs.

Verificarlo computes the number of significant bits to evaluate the numerical accuracy of a computation. It captures the number of accurate bits in the FP mantissa against a chosen reference.  Unfortunately, an exact reference value is not known beforehand for many complex programs or intermediate computations. To overcome this problem, Verificarlo uses Monte Carlo arithmetic (MCA)~\cite{PARKER1997}, a stochastic method that can simulate numerical errors and estimate the number of significant bits directly: the result of each FP operation is replaced by a perturbed computation modeling the losses of accuracy.
From a set of MCA samples, it is possible to estimate the significant bits of a computation,
$s_{2} = -\log_{2}|{\sigma}/{\mu}| $,
where $\sigma$ and $\mu$ are the sample's standard deviation and mean~\cite{confidence_intervals}.

Verificarlo includes six backends, which are extensively documented in the user manual. The two most important backends are:
the \texttt{MCA} backend, described previously, and the \texttt{VPREC} backend that emulates the effect of using mixed-precision in a program~\cite{Chatelain2019automatic}.

Verificarlo CI automates numerical accuracy tests by using a separate Git branch to store test results. Whenever users make modifications to the main branch, a remote runner carries out predefined tests and uploads the results to the CI branch. Users can view their results dynamically using a simple web server.

Verificarlo CI offers a command-line interface that helps configuring the CI pipeline on a given application and hooks it up to a GitLab or GitHub repository : it automatizes the initial setup by creating both the CI pipeline and the dedicated results branch. Users are then free to further customize their pipelines.

Developers use a dedicated C or Fortran API to include \emph{probes} in their code. Each probe is associated with a test and a variable name. During testing, the probe measures the accuracy of a chosen variable. Optionally, an alert can be triggered if the relative or absolute error exceeds a user-set threshold. In the below C example, the probe is identified by the \lstinline{"test"}/\lstinline{"var"} couple, and an absolute error threshold is set to $0.01$:

\begin{lstlisting}
vfc_probe_check(probes, "test", "var", var, 0.01);
\end{lstlisting}

In order to be able to run the tests, Verificarlo CI  requires a description of the tests and backends to run. It is specified in a JSON configuration file which supports complex test setups. The test results are exported to an HDF5 file, a hierarchical format commonly used in HPC applications.  Test results are stored on the dedicated CI branch, allowing robust archival. The HDF5 files can optionally embed the raw test results. In the default CI workflow, this raw data is stored as a job artifact and accessible for a limited time, to enable user defined additional analysis.
Finally, Verificarlo CI analyzes the data and generates dynamic reports organized into different views: temporal, cross-test, or cross-variable comparisons and accuracy violations.

\section{Mixed-precision for Nekbone proxy application}
Nek5000\footnote{\url{https://nek5000.mcs.anl.gov/} and \url{https://github.com/Nek5000/Nekbone}} is a high-order solver for Computational Fluid Dynamics (CFD) based on the Spectral Element Method (SEM) that solves the Navier-Stokes equation for incompressible \begin{wrapfigure}{r}{0.45\textwidth}
    \centering
    \vspace*{-2.mm}
    \resizebox{0.48\textwidth}{!}
    {
        \input{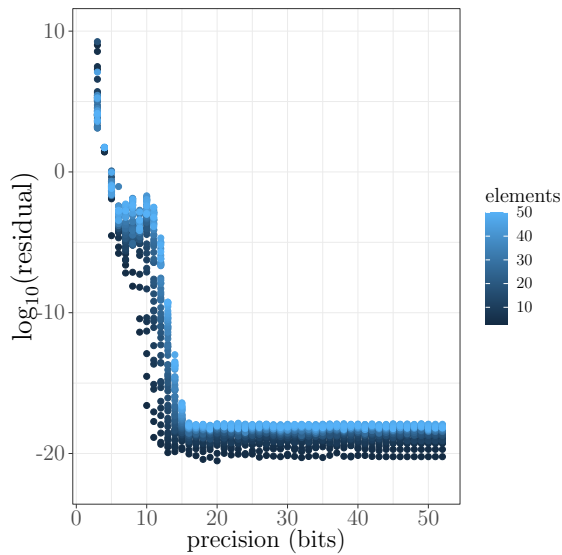}
    }
    \caption{Examining precision needs in Nekbone for various numbers of elements: the residual
        ($L^2$ norm) in CG.
        \label{fig:nekbone}}
    \vspace*{-7.5mm}
\end{wrapfigure}
flow. Nekbone is a proxy application for Nek5000 that focuses on important computational and scaling aspects of the entire solver.
We used the \texttt{VPREC} backend to examine precision appetites in Nekbone
using the polynomial degree of 10 and different number of elements. The results
of tracking the residual of the Conjugate Gradient (CG) solver,
see Figure~\ref{fig:nekbone}, suggest a possibility of using as little as 16 bits of
mantissa (the beginning of the plateau) and still being able to converge, while
the original version relies on FP64 (double precision) with 52 bits of mantissa.
We verified this assumption with the \texttt{MCA} backend, confirming such a
possibility for the precision reduction to single in the CG solver on CPUs.
Recently, following this precision inspection, we modified Nekbone to support
mixed single-double precision and  we were able to reduce the time-to-solution
by $34\,\%$.
Once a suitable precision is found, a Verificarlo CI probe can be inserted in the code, to automatically monitor the residual error of each subsequent code version.

\section{Tracking accuracy in the QMCkl library}

The Sherman-Morrison-Woodbury (SMWB) kernel was developed as part of the QMCkl library~\footnote{\url{https://github.com/TREX-CoE/qmckl}}, an open-source library of highly-optimized building blocks for implementing Quantum Monte Carlo methods in the TREX European Center of Excellence.

Given a matrix $A$ and its inverse $A^{-1}$, Sherman-Morrison (SM) is a formula to efficiently compute the inverse after a rank-1 update $uv^T$ on $A$:

\begin{equation}
    (A + uv^T)^{-1} = A^{-1} - \frac{A^{-1} uv^T A^{-1}}{1 + v^T A^{-1} u}
\end{equation}

This formula can be generalized for rank-$k$ updates using the Woodbury (WB) formulation~\cite{woodbury1950inverting}.
In WB, the denominator of SM is replaced by the inverse of a small square $k\times k$ matrix. For $k=2$ and $k=3$, WB is expected to be faster than iterating 2 or 3 times with SM.
QMCkl  implements different algorithms to apply SM with a set of updates $(u_j, v_j)$, for $j = 1, ..., m$. The naive approach, \textit{SM1}, applies these updates in sequence.

Depending on the updates ordering, the SM denominator can be close to zero, meaning that the matrix $A$ becomes singular. This makes the method numerically unstable.  A refined algorithm using Slagel splitting~\cite{Slagel2015TheSM} is called \textit{SM2}. Below a minimum threshold for the denominator, the update is split in two, the first half is applied, and the second half enqueued with remaining updates.

To implement the Woodbury formula, blocks of rank-3 and rank-2 updates are built. If the intermediate matrix update is singular, the corresponding updates are split with SM2. This method is called \textit{SMWB}.
Since \textit{SMWB} changes the order of operations, one must ensure that the numerical accuracy is preserved compared to \textit{SM2}.

\begin{figure}
    \begin{minipage}{0.47\textwidth}
        \label{boxplot}
        \begin{center}

            \begin{tikzpicture}

                \begin{axis}[
                        legend to name={legend},
                        boxplot/draw direction=y,
                        ylabel=$s_2$,
                        xlabel=,
                        xtick={1,2, 3},
                        xticklabels={SM1,SM2,SMWB},
                        scale=0.6
                    ]

                    \addplot+[boxplot] table [
                            col sep=comma,
                            x expr=\coordindex,
                            y=frob2_sm1_s2]
                        {boxplot_6f282f3_raw.csv};

                    \addplot+[boxplot] table [
                            col sep=comma,
                            x expr=\coordindex,
                            y=frob2_sm2_s2]
                        {boxplot_6f282f3_raw.csv};

                    \addplot+[boxplot] table [
                            col sep=comma,
                            x expr=\coordindex,
                            y=frob2_smwb1_s2]
                        {boxplot_6f282f3_raw.csv};

                    \addplot[mark=none] coordinates {(3,33.2)} node[pin=235:{dataset 4263}]{} ;
                \end{axis}

            \end{tikzpicture}

        \end{center}
        \caption{Significant bits of Frobenius norm, for all datasets and algorithm combinations, for commit \lstinline{6f282}, grouped by algorithms. SMWB fails catastrophically in some cases.\label{boxplot}}
    \end{minipage}\hspace{3mm}%
    \begin{minipage}{0.5\textwidth}

        \pgfplotstableread[col sep=comma]{cycle_04263_frob2.csv}\datatable
        \begin{tikzpicture}
            \begin{axis}[
                    xtick=data,
                    xticklabels from table={\datatable}{commit},
                    ymin=0,
                    width=0.5\textwidth,
                    height=0.4\textwidth,
                    x tick label style={
                            rotate=45,
                            anchor=east,
                        },
                    xlabel=commits ,
                    ylabel=$s_2$,
                    x label style={at={(axis description cs:0.5,-0.2)}},
                    legend style={at={(0.75,0.53)},anchor=north},
                    scale=2.2
                ]
                \addplot table [col sep=comma, x expr=\coordindex, y=sm1_frob2_s2]{cycle_04263_frob2.csv};
                \addlegendentry{SM1}
                \addplot table [col sep=comma, x expr=\coordindex, y=sm2_frob2_s2] {cycle_04263_frob2.csv};
                \addlegendentry{SM2}
                \addplot table [col sep=comma, x expr=\coordindex, y=smwb_frob2_s2]{cycle_04263_frob2.csv};
                \addlegendentry{SMWB}

            \end{axis}
        \end{tikzpicture}
        \caption{Significant bits of Frobenius norm, for our different algorithms, over commits for dataset 4263. SMWB's accuracy improves after the fix.\label{slagel_plot}}
    \end{minipage}%
\end{figure}

To track the accuracy of these algorithms during development with Verificarlo CI, we use a large number of datasets from a QMC=Chem~\cite{scemama2012qmc} use case on Benzene. All datasets are run with all algorithms in \texttt{MCA} mode. The main development branch in the repository was instrumented with probes identified with \textit{dataset number} / \textit{algorithm} couples allowing a factored analysis in the dynamic reports. Finally, we set up a CI branch using the command-line helper to track accuracy on the main development branch, from which Figures \ref{boxplot} and \ref{slagel_plot} were generated.

During the development of SMWB, the \emph{run inspection} report, reproduced in Figure~\ref{boxplot}, highlighted some outputs for which SMWB fails with a high error under the MCA backend. After investigation, we discovered that in the initial implementation of SMWB, delayed updates were directly applied after each WB step. This reduces the numerical stability because it increases the probability of producing singular intermediate matrices. It was fixed in commit \lstinline{67f53} by moving all the updates to the very end of the update queue as shown in Figure \ref{slagel_plot}, obtained from the \emph{temporal view}.

\section{Conclusion}
\label{}
Verificarlo CI automates numerical accuracy tests within a continuous integration workflow: it grants users the ability to define such tests. It provides an easy way to visualize results throughout the development process of a code. Better integration of numerical checks in the CI process saves developers precious time to focus on their area of expertise.

Verificarlo CI has been used in the context of TREX and CEEC EuroHPC JU
Centers of Excellence to detect numerical regressions, pin-point faulty commits, and predict the effect of mixed-precision.
A tutorial demonstrating its use is available at \url{https://github.com/verificarlo/vfc_ci_tutorial}. Furthermore, we believe that using such a tool as a part of the regular CI/ CD process would help for early stage identification of numerical bugs and re-ensuring numerical reliability of codes.

    {
        \textbf{Acknowledgements}
        This work was partially supported from EC by EuroHPC Centers of Excellence TREX (952165) and CEEC (101093393), as well as by the French National Agency for Research via the InterFLOP project (ANR-20-CE46-0009).
    }

\bibliography{bibliography}{}
\bibliographystyle{plain}

\end{document}